\documentclass[preprint,authoryear,12pt]{elsarticle}

\usepackage{graphics}

\usepackage{amssymb}
\usepackage{hyphenat}
\journal{Advances in Space Research}

\newcommand{\be}{\begin{equation}}
\newcommand{\ee}{\end{equation}}
\newcommand{\beq}{\begin{eqnarray}}
\newcommand{\eeq}{\end{eqnarray}}

\newcommand{\adv}{    {\it Adv. Space Res.~}}
\newcommand{\annG} {{\it Annales Geophysicae~}}

\newcommand{\ag}{     {\it Ann. Geophys.~}}

\newcommand{\grl}{    {\it Geophys. Res. Lett.~}}

\newcommand{\jgr}{    {\it J. Geophys. Res.~}}
\newcommand{\mnras}{  {\it Mon. Not. Roy. Astron. Soc.~}}
\newcommand{\na}  {{\it New Astronomy ~}}
\newcommand{\nat}{    {\it Nature~}}

\newcommand{\solphys}{{\it Solar Phys.~}}

\newcommand{\ssr}{    {\it Space Sci. Rev.~}}

\newcommand{\cp}{\citep} 
\newcommand{\ct}{\citet} 
\newcommand{\DST}{$\rm D_{ST}$} %
\newcommand{\DG}{$\rm ^o$}
\begin{document}

\begin{frontmatter}

\title{A Wavelet Based Approach to Solar--Terrestrial Coupling}

\author{Ch. Katsavrias$^{1}$}
\cortext[cor]{Corresponding author}
\fntext[footnote2]{Tel. 00302107276855}
\ead{ckatsavrias@phys.uoa.gr}

\author{A. Hillaris$^{1}$}
\author{P. Preka--Papadema$^{1}$}

\address{$^{1}$Department of Astrophysics, Astronomy and Mechanics, Faculty of Physics, University of Athens, Panepistimiopolis Zografos (Athens) , GR--15783, Greece}

\begin{abstract}
{Transient and recurrent solar activity drive geomagnetic disturbances; these are quantified (amongst others) by \DST, AE indices time-series. Transient disturbances are related to the Interplanetary Coronal Mass Ejections (ICMEs) while recurrent disturbances are related to corotating interaction regions (CIR).  We study the relationship of the geomagnetic disturbances to the solar wind drivers within solar Cycle 23  where  the  drivers are represented by ICMEs and CIRs occurrence rate and compared to the \DST~and AE as follows: terms with common periodicity in both the geomagnetic disturbances and the solar drivers are, firstly, detected using continuous wavelet transform (CWT). Then,  common power and phase coherence of these periodic terms are calculated from the cross-wavelet spectra (XWT) and wavelet-coherence (WTC) respectively. In time-scales of \mbox{$\approx 27$ days} our results indicate an anti-correlation of the effects of ICMEs and CIRs on the geomagnetic disturbances. The former modulates the \DST~ and AE time series during the cycle maximum the latter during periods of reduced solar activity.   The phase relationship of these modulation is highly non-linear. Only the annual frequency component of the ICMEs is phase--locked with \DST ~and AE. In time-scales of \mbox{$\approx$1.3-1.7} years the CIR seem to be the dominant driver for both geomagnetic indices throughout the whole solar cycle 23.}
\end{abstract}


\end{frontmatter}

\parindent=0.5 cm

\section{Introduction}\label{Introduction}
The connection of solar activity to geomagnetic disturbances, dubbed Solar--Terrestrial Coupling, remains an open field of research. The effects on Earth appear as geomagnetic disturbances driven by the solar wind--magnetosphere interaction and quantified by geomagnetic indices \citep[see review by~][]{Akasofu2011}. 

\citet{Feynman1982} and \citet{Du2011a}, indicated that  the annual values of the geomagnetic index \textit{aa} could be the resultant of two components: one originating from solar transient  (or sporadic) activity and in phase with the solar cycle; the other was related to recurrent solar drivers with peak in the declining phase \citep[see also~][]{Richardson2012}. Along the same line \citet{Cliver1995}~provides a historical review of the solar-terrestrial research since 1930, and the two basic types of geomagnetic storms: recurrent and sporadic. The studies, mentioned above, propose two classes of geomagnetic--solar drivers on a time scale of approximately a year as \citet{Feynman1982} and \citet{Du2011a} used annual averages of \textit{aa}~in their study. The interplanetary coronal mass ejection (ICME) is the major driver of transient geomagnetic activity.The solar recurrent activity, on the other hand, is driven by High Speed Solar Wind Streams (HSSWS) and Co-rotating Interaction Regions (CIR)  \citep{Schwenn2006,Pulkkinen2007}. \citet{Borovsky2006b} and \citet{Richardson2012} indicate, also, that the different driver classes (CIR, ICME) result in distinct geomagnetic  disturbances; the ICMEs, for example,  induce higher ring current, manifested by a high negative peak in \DST. 

The solar--geomagnetic coupling, when studied in the frequency plane manifests itself with periodic terms having  the same periodicity in the solar drivers and the geomagnetic indices time series. The basic periodicity is the 11/22 year solar cycle (sunspot and magnetic respectively), yet quasi-periodic variations on shorter time-scales have been reported.   

\citet{Lou2003} found A$_p$ index periodicities of 187, 273 and 364 days in the 1999--2003 time interval. Periodicities of about 27.5, 13.5, 9.1, and 6.8 days, due to the solar rotation have been identified in the solar wind speed and the IMF polarity \citep{Gonzalez1987, Gonzalez1993, Svalgaard1975, Fenimore1978,Sabbah2011}. \citet{Kudela2010} reported that a range of periodicities, \mbox{1.7-2.2} years, appear in cosmic rays during the time interval 1951--2010, while \citet{Mavromichalaki2003} published similar results for the \mbox{1953-1996} interval. \citet{VGalicia1996, Mursula1999} and \citet{Nayar2002} reported different periodic variations of the geomagnetic activity index A$_p$; \mbox{1.3-1.4} years during {\em{even}} cycles and of \mbox{1.5-1.7} years during {\em{odd}} ones. 

\citet{Katsavrias2012} examined the \mbox{1966-2010} time period for periodicity in the solar activity, the solar wind speed, interplanetary magnetic field and the geomagnetic indices ~using wavelet analysis. Within the examined time-series time-localized common spectral peaks, between the fluctuations in the solar wind characteristics and the geomagnetic indices were detected. Certain periodicities were dominant within specified intervals which, at times, were different for different geomagnetic indices.

The interdependence between different time series requires a different wavelet based approach. In this case cross wavelet transform and wavelet coherence \cp[XWT and WTC~][]{Grinsted2004} are used for the quantification of the interdependence. This approach has been, already, used in the study of common periodicities between two time--series and the corresponding phase relationship between them. \ct{Galicia2008} studied the coherence of the sunspots with open solar magnetic fluxes. \ct{Deng2012} investigated the coronal index--sunspot numbers phase relationship finding coherent behaviour in low--frequency components corresponding to the 11--year Schwabe cycle; this coherence was absent in the high--frequency components. \ct{Deng2013} applied this method between 10.7 cm solar radio flux and sunspot numbers from 1947 February to 2012 June; the phase relationship between the time series was found both time and frequency dependent.
\begin{figure*} 
\centering 
\includegraphics[trim=0.0cm 0.0cm 0.0cm 0.0cm,clip,width=1.0\textwidth,height=0.40\textheight]{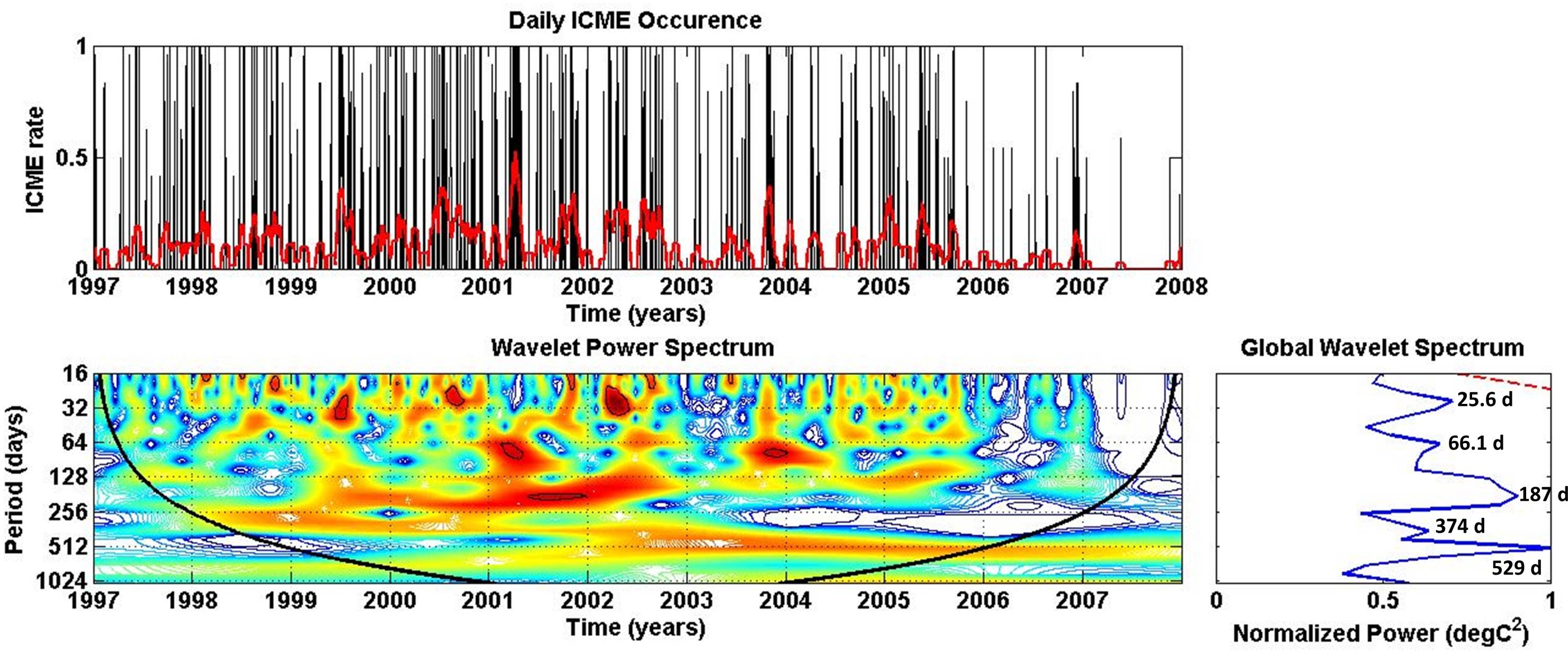} 
\caption{Time-series (upper panel), Wavelet power (lower panel, left) and global wavelet spectra (lower panel, right) of ICMEs occurrence; the red line is the 27-days moving average smoothed time-series. The Wavelet power display is colour-coded with red corresponding to the maxima; the black contour is the cone of influence of the spectra, where edge effects in the processing become important. The dashed line in the global spectra represent a confidence level above 95 \%.}
\label{F1}
\end{figure*}

In this work a refinement of the \citet{Katsavrias2012} wavelet based approach is presented which aims at the detection of common and coherent periodicity and phase relationship between the ICMEs, CIRs and the \DST, AE geomagnetic indices time-series by means of cross wavelet transform and wavelet coherence calculations. 

\begin{figure*} 
\centering 
\includegraphics[trim=0.0cm 0.0cm 0.0cm 0.0cm,clip,width=1.0\textwidth,height=0.40\textheight]{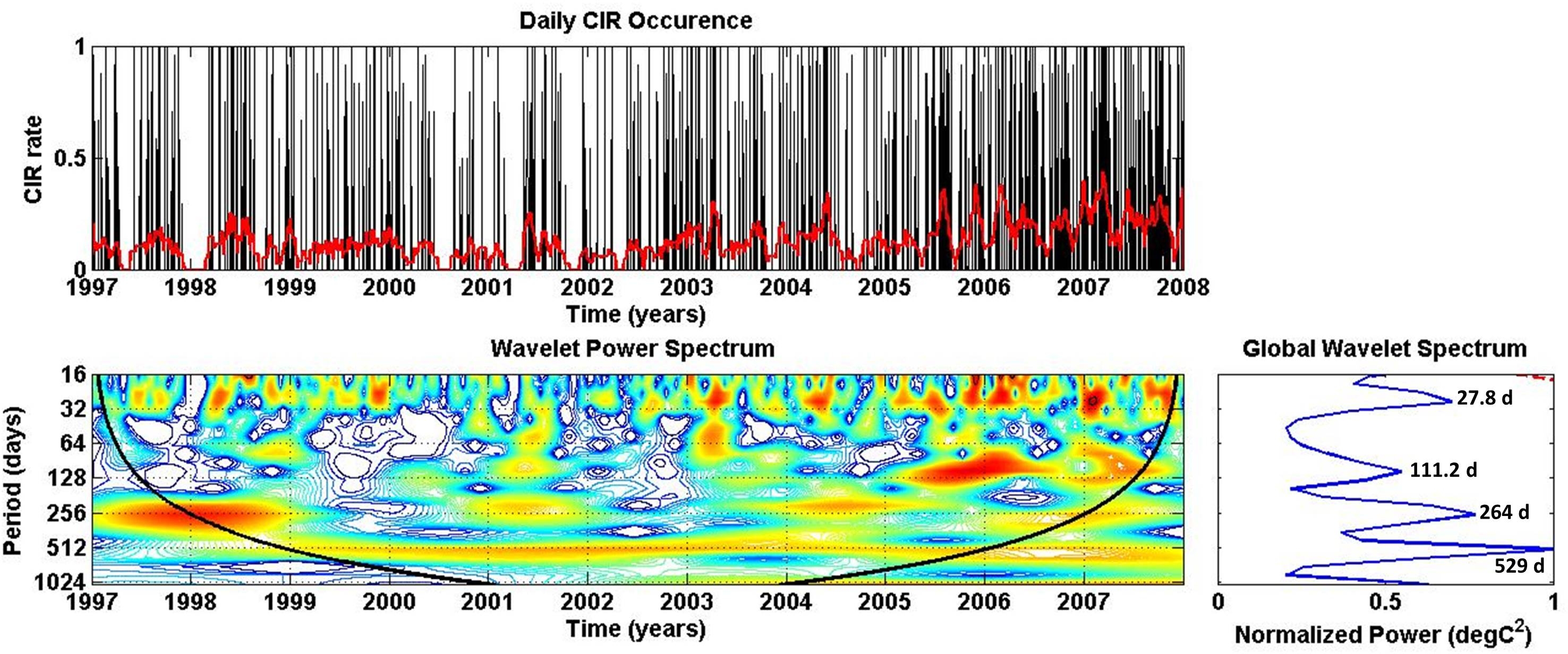} 
\caption{Same with figure \ref{F1} but for CIR occurrence.}
\label{CIR}
\end{figure*}
\section{Data Selection} \label{Obs}
We used time-series of the occurrence rate of the geomagnetic drivers, ICMEs, CIR and of different geomagnetic indices, representative of the conditions in the magnetosphere, as follows:
\begin{itemize}
\item {ICMEs per day from the \citet{Jian2006a} catalogue on line\footnote{http://www$-$ssc.igpp.ucla.edu/\~jlan/STEREO/Level3/STEREO\_Level3\_ICME.pdf recent updates of the catalogue extend beyond 2006.}.  The daily rate is the duration of the ICME passage on that day, in hours, divided by 24. Two more ICME lists by \citet{Richardson2010} and \citet{Mitsakou2014} were available yet the selection does not affect our analysis as the three lists differ little from each other and exhibit the same trends in the ICME occurrence rate \citep{Mitsakou2014}.}
\item{CIRs  per day from the \citet{Jian2006b}  list on line\footnote{www$-$ssc.igpp.ucla.edu/\~jlan/STEREO/Level3/STEREO\_Level3\_SIR.xls}. The daily rate is the duration of the CIR passage on that day, in hours, divided by 24 defined similarly to the ICME rate in the previous bullet. We selected CIRs because their geomagnetic effectiveness is greater, on average, than the other stream interaction regions. }
\item {Geomagnetic indices from the OMNIweb database: The \DST, represents the strength of the Earth ring current; values below -30 nT indicate a geomagnetic storm. The AE quantifies sub-storms as it represents auroral electrojet intensity ~\citep{Mayaud1980}.}

\end{itemize}
\noindent Our data-set covers solar cycle 23, from January 1st, 1997 to December 31st, 2007, and consists of daily average values. 

\section{Results and Discussion}\label{Res}

\begin{figure*} 
\centering 
\includegraphics[trim=0.0cm 0.0cm 0.0cm 0.0cm,clip,width=1.0\textwidth,height=0.95\textheight]{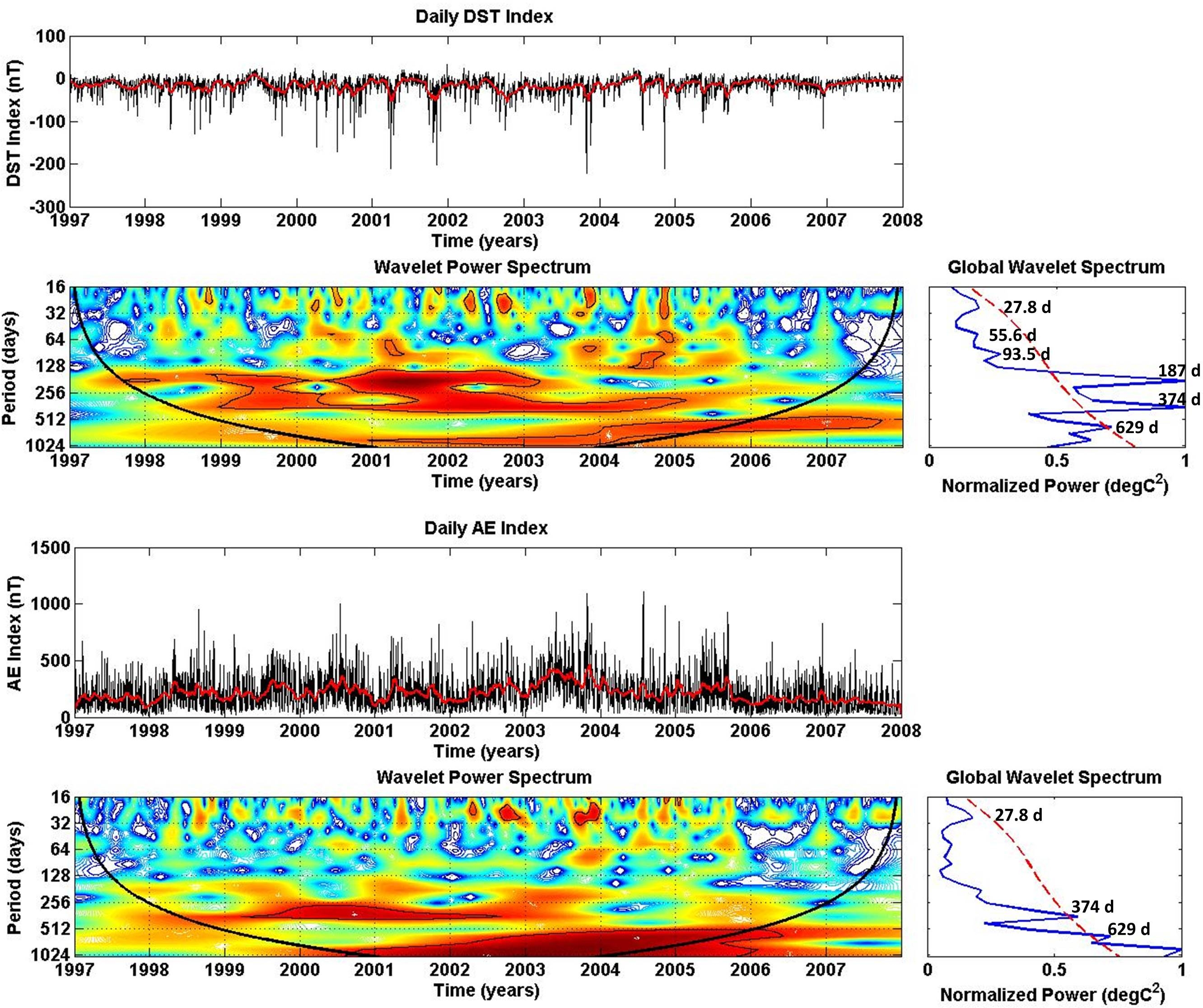} 
\caption{Same with figures \ref{F1} and \ref{CIR} but for geomagnetic indices (top to bottom): \DST ~and AE.}
\label{F3}
\end{figure*}

\subsection{Wavelet analysis}\label{wavelet}
The analysis of a function in time, be it F(t), into an orthonormal basis of \textit{wavelets} is conceptually similar to the Fourier Transform. The latter however is localised in frequency (or time--scale) only while the former, being localised in frequency and time, allows the local decomposition of Non--stationary time series; a compact, two dimentional, representation may be thus obtained \citep[see~][]{Morlet1982,Torrence1998a}. The \textit{wavelets} forming the basis are derived from an integrable zero-mean \textit{mother wavelet} $\psi$(t) and the wavelet transform of F(t), be it W(t,f), is caculated as the convolution of this function with the \textit{mother wavelet} duly shifted and scaled in time {\rm$\psi(f\cdot(\tau-t))$}:
\be
{\rm{W(t,f) = }}\int_{ - \infty }^{ + \infty } {{\rm{F(\tau )}}\sqrt {\rm{f}} } {\rm{\psi }}^{\rm{*}} \left( {{\rm{f}}\left( {{\rm{\tau  - t}}} \right)} \right){\rm{d\tau }}
\ee
\noindent{where * denotes complex conjugate, the scale factor $f$ represents frequency and $\sqrt{f}$ is necessary to satisfy the normalization condition; the wavelet transform represents a mapping of F(t) on the t-f plane. 

The \textit{mother wavelet} which in our case is the {\em{Morlet wavelet}} which consists of a plane wave modulated by a Gaussian: ${\rm{\psi }}_{\rm{n}} {\rm{(f) = \pi }}^{{{\rm{1}} \mathord{\left/{\vphantom {{\rm{1}} {\rm{4}}}} \right. \kern-\nulldelimiterspace} {\rm{4}}}} {\rm{exp(i\omega n)}} \cdot {\rm{exp}}\left( {{\rm{ - n}}^{\rm{2}} {\rm{/2}}} \right)$; and {\rm$\omega$} is a constant \citep[usually set to 6, see~][]{Torrence1998a}. This type of  \textit{mother wavelet} is quite common in astrophysical signals analysis facilitating comparison with previously published works. Due to its Gaussian support, the Morlet {\em{wavelet}} expansion inherits optimality as regards the uncertainty principle \citep{Morlet1982}.

The average of the wavelet power spectral density $\left\| {{\rm{W(t,f)}}} \right\|^2$ on time (t) is the global wavelet spectrum \citep[see~][]{Torrence1998a} and is given by:
\be 
 \overline {{\rm{W}}\left( {\rm{f}} \right)} {\rm{ = }}\frac{{\rm{1}}}{{\rm{N}}}\sum\nolimits_{{\rm{n = 1}}}^N {\left\| {{\rm{W}}_{\rm{n}} {\rm{(f)}}} \right\|} ^2  \\ 
\ee
\noindent for discrete time. The global wavelet spectrum is an unbiased and consistent estimation of the true power spectrum of a time series and generally exhibits similar features and shape as the corresponding Fourier spectrum. }

\subsection{Periodicities in Geomagnetic Indices and Solar Wind Drivers}

From the wavelet power spectra and the global wavelet spectra,  presented in \ref{wavelet}, we identified periodic components of the time-series in \ref{Obs}, in the range from days to a year, within a confidence level\footnote{The confidence level is defined as the probability that the true wavelet power at a certain time and scale lies within a certain interval around the estimated wavelet power.} of 95\%. We should note, at this point, that due to the statistical nature of the methodology in use, and the dependence of the results on sample size, we have retained some global peaks that are a little below the 95\% threshold but persist for long periods of time in the power spectrum.  The periodicities in the time series appear to dominate certain time intervals being absent from others.

In figures \ref{F1}--\ref{F3} we identify short (close to the solar rotation) and mid-term (more than 3 months) periodicities of varying power, localized in time. In each of these figures we present the time-series (top panel) to be analysed, the wavelet power spectrum which depicts the time localized periodicities (bottom panel) and the global wavelet spectrum (bottom right panel); the latter is the average over time of each periodic component and facilitates the identification of the peak of each range of periodicities. The studied time--series are described below:
\begin{itemize}
\item{ ICME rate (Figure \ref{F1}): The mid-term CME periodicities at 187 and 374 days (approximately six and twelve months) at the solar cycle 23 maximum \citep{Polygiannakis2002, Lou2003}, are present in the ICME rate time series under the 95\% confidence level; a prominent peak of \mbox{$\sim$187} days appears only during 2001. Moreover a peak at 66 days appears during 2001 and late 2003. Sporadic short-term periodicities, peak at approximately 25 days, are also present, around the solar maximum (1999--2002) yet they are mostly below the 95\% confidence level of the global spectrum. }

\item{ CIR occurrence rate (Figure \ref{CIR}): Ephemeral periodicities (peak at 27.8 days) are most pronounced during the decline phase of the solar cycle 23 yet generally below the 95\% confidence level. The global wavelet spectrum also shows mid-term periodicities with peaks at \mbox{$\sim$111} and \mbox{$\sim$264} days which appear during the decline and rising phase respectively but with power levels under the 95\% confidence limit.}

\item{ Geomagnetic Indices (Figure \ref{F3}): Both indices time-series have intermittent, short-term, sporadic, low--confidence (mostly less than 95\%) periodic components mostly during the late solar maximum and the decline phase \mbox{(2002-2004).} The \DST ~exhibits a pronounced, 374 days, peak (annual periodicity, confidence level exceeds 95\%)  in 1999--2004 and a second, 187 days  (semi annual periodicity), peak in 1998--2003.  On the other hand, AE ~exhibits only the annual periodicity in the \mbox{1999-2002} interval.}
\end{itemize}

The periodic terms common to two or more time-series were analyzed further in the following subsections, using XWT and WTC. 

\subsection{Cross Wavelet  Analysis and Wavelet Coherence}\label{XWT}
The \textit{Cross Wavelet Transform} (XWT) makes use of the wavelet analysis in the examination of causal relationships in time frequency space between two time series X and Y with corresponding CWTs: ${\rm W_n^X(f)}$ and ${\rm W_n^Y(f)}$. The cross-wavelet transform of the time-series X and Y is defined as: ${\rm W_n^{XY}(f)= W_n^X(f)\cdot W_n^Y(f)^*}$, with * denoting complex conjugate. 

The result is, in general, complex; the modulus, ${\left\|\rm W_n^{XY}\right\|}$, indicates \textit{regions} in the (t-f) space of high common power and the phase, arg(${\rm W_n^{XY}}$), of the XWT represents relative phase relationship of the time-series to be compared:
\be 
{\rm
{\tan ^{ - 1}}[\frac{{{\mathop{\rm Im}\nolimits} (\left| {W_n^{XY}(s)} \right|)}}{{{\mathop{\rm Re}\nolimits} (\left| {W_n^{XY}(s)} \right|)}}]
}
\label{A4}
\ee
\noindent Said regions of high common power and consistent phase relationship suggest causal relationship between X and Y. From the phase of the XWT a measure of \textit{Wavelet Coherence} (WTC) between ${\rm W_n^X}$ and ${\rm W_n^Y}$ will be derived below. The statistical significance of the Cross Wavelet Spectrum was estimated following \ct{Torrence1998a} and \ct{grinsted2006}. 

The Cross-Wavelet Transform is used in the calculation of the degree of  {\emph{cause and effect}}  dependence of the  {\emph{geomagnetic response}} to the Solar Activity and the Solar Wind as all of them are represented by time series. 

The Wavelet Coherence (WTC) is an estimator of the confidence level for each detection of a time--space region of high common power and consistent phase relationship, calculated by the Cross Wavelet Transform, between two time-series. The measure of wavelet coherence is defined between two continuous wavelet transforms and it may indicate coherence with high confidence level even though the common power is low; it closely resembles a localized correlation coefficient in time--frequency space and varies between 0 and 1. It is used alongside the  Cross Wavelet Transform as the latter appears to be unsuitable for significance testing the interrelation between two processes \citep{Maraun2004}. Following \citet{Torrence1998b} we define the wavelet coherence of two time series, let them be X and Y:
\be
{\rm
R_n^2(f) = \frac{{{{\left| {S({f^{ - 1}}W_n^{XY}(f))} \right|}^2}}}{{S({f^{ - 1}}{{\left| {W_n^X(f)} \right|}^2}) \cdot S({f^{ - 1}}{{\left| {W_n^Y(f)} \right|}^2})}}
}
\label{A8}
\ee
\noindent where S is a smoothing operator. As this definition closely resembles that of a traditional correlation coefficient,  we might think the wavelet coherence as a correlation coefficient localized in  time frequency space. The statistical significance level of the wavelet coherence is estimated using Monte Carlo methods. 

A detailed description of the Wavelet-Based Method for the Comparison of Time Series may be found in \citet{Torrence1998b,Grinsted2004,grinsted2006}.  The Matlab package of the National Oceanography Centre, Liverpool, UK\footnote{http://noc.ac.uk/usingscience/crosswaveletwaveletcoherence} was used in the calculation of the WXT and WTC. 
\begin{figure*} 
\centering 
\includegraphics[trim=0.0cm 0.0cm 0.0cm 0.0cm,clip,height=0.80\textheight,width=1.2\textwidth]{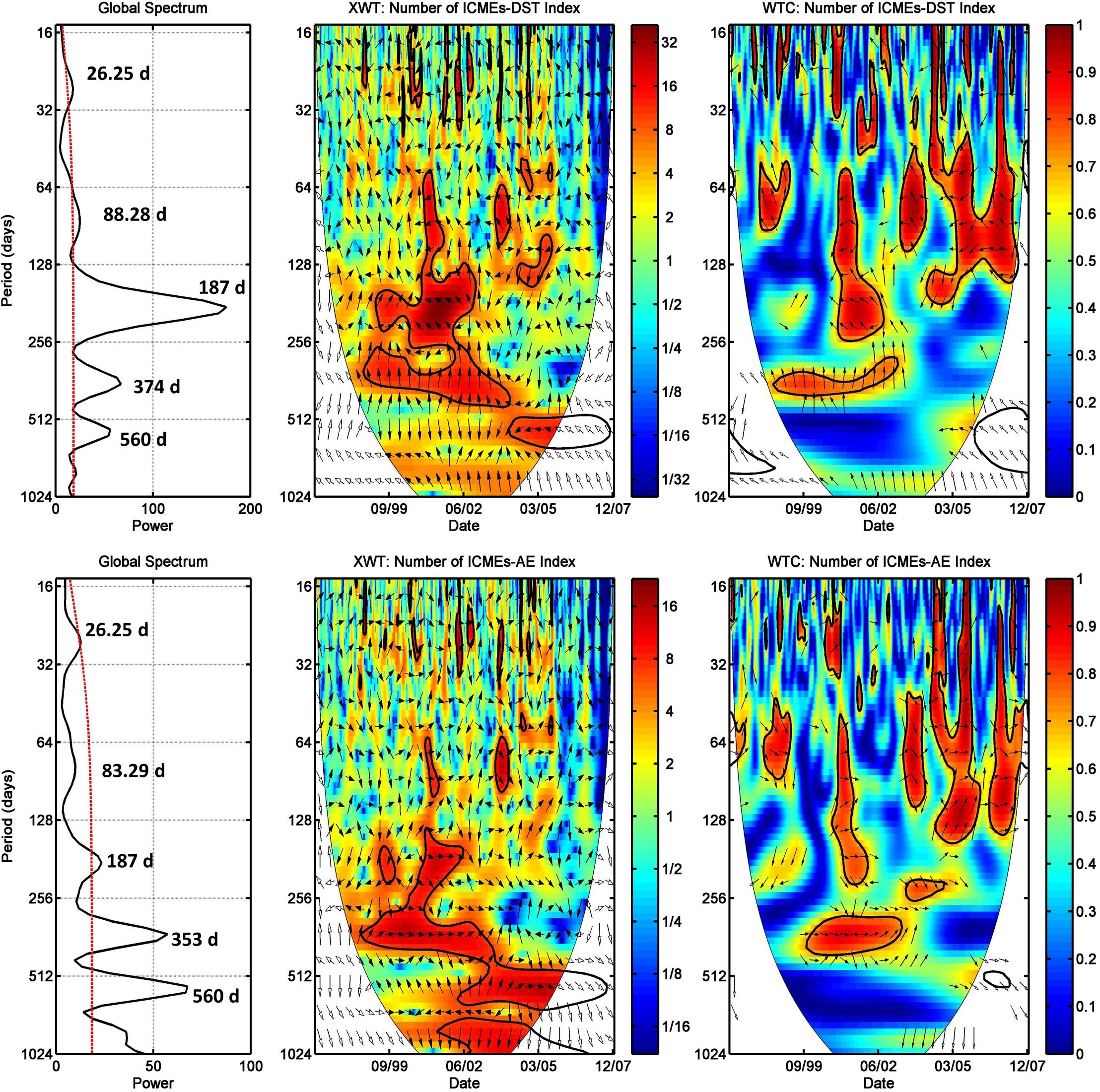} 
\caption{Global wavelet (left), Cross-wavelet transformation (XWT, middle) and Wavelet Coherence (WTC, right) of the ICME occurrence rate and the geomagnetic indices; The dashed red line corresponds to the 95\% confidence level of the global wavelet. The thick black contours mark the 95\% confidence level, and the the thin line indicates the cone of influence (COI). The colour-bar of the XWT indicates the power of period range; the colour-bar of the WTC corresponds to the significance level of the Monte-Carlo test. The arrows point to the phase relationship of the two data series in time-frequency space: (1) arrows pointing to the right show in-phase behavior; (2) arrows pointing to the left indicate anti--phase behavior; (3) arrows pointing downward indicate that the first dataset is leading the second by 90\DG}
\label{ICME_Indices}
\end{figure*}
\subsection{ICMEs--Geomagnetic Effects Relationship}\label{ICMEGeo}
{ Figure \ref{ICME_Indices}, shows the cross--wavelet transform and wavelet coherence calculations used to study the interrelation of ICMEs and geomagnetic indices. The middle panels show the cross-wavelet spectrum (XWT, see \ref{XWT}) of the two time series under examination; the common power of the time-series pair is colour coded in the time-period domain. The left panels depict the time average of the XWT spectrum and the right panels the wavelet coherence. The latter is  the correlation coefficient (Equation \ref{A8}) of the time-series wavelet transform phase. Arrows indicate the phase relationship between the two data series  (Equation \ref{A4}), in time-frequency space: Those pointing to the right correspond to  in--phase behavior those to the left  anti--phase. The downwards pointing arrows indicate 90\DG~ lead of  the first data-set. Since geomagnetic strorms imply large negative values of \DST~ the convention is reversed and now left indicates in phase and downwards pointing arrows imply that the ICME time series leads the \DST. The same reversed convention holds in section \ref{CIRGeo} for the CIR-\DST ~time series}.

Similar to \citet{Katsavrias2012}, we consider significant, in XWT, WTC and in continuous wavelet, the shared periodicities which persist for an interval of at least 4--5 times its period and with a coherence coefficient above 0.8.

We discuss below the approximately 27 days, 3 month, semi-annual, annual and 560 days periodicity. Those represent the pronounced peaks of the geomagnetic disturbances and their drivers, cross wavelet transform (XWT):

\begin{itemize}
\item{ ICMEs--\DST~index: The ICME rate and the \DST~index share the annual periodicity (peak at the 374 days)  most of the solar cycle 23 (1998--2003) with  generally phase-locked behavior; the ICME rate leads by 90\DG~ the \DST. The first harmonic (187 days peak), appears in 1999--2002 with varying phase behavior.  Short-term 26 days periodicity appears intermittently in short intervals throughout the cycle; prominent peaks appear 1999, 2001 and 2005 with varying phase ICME--\DST~behaviour. A prominent peak at approximately 88 days is also present at 2001 and late 2003.}

\item{ ICMEs--AE index: The ICME rate and the AE index time-series share the, approximately, annual and semi-annual periodicities (peaks at 353 and 187 days) in 1999--2002 yet only the former exhibits in-phase relationship. Short-term,  27 days, periodicities appear sporadically  throughout the cycle 23 with varying phase behavior. The approximately 3 months periodicity (peak at 83 days) is also present but mostly below the 95\% confidence level. }
\end{itemize}
\noindent The 560 days periodicity appears in both ICME--\DST~ and ICME--AE XWT during the whole solar cycle yet is above the 95\% confidence level only during 2004--2005 and 2002--2005 time intervals respectively.

\begin{figure*} 
\centering 

\includegraphics[trim=0.0cm 0.0cm 0.0cm 0.0cm,clip,height=0.80\textheight,width=1.2\textwidth]{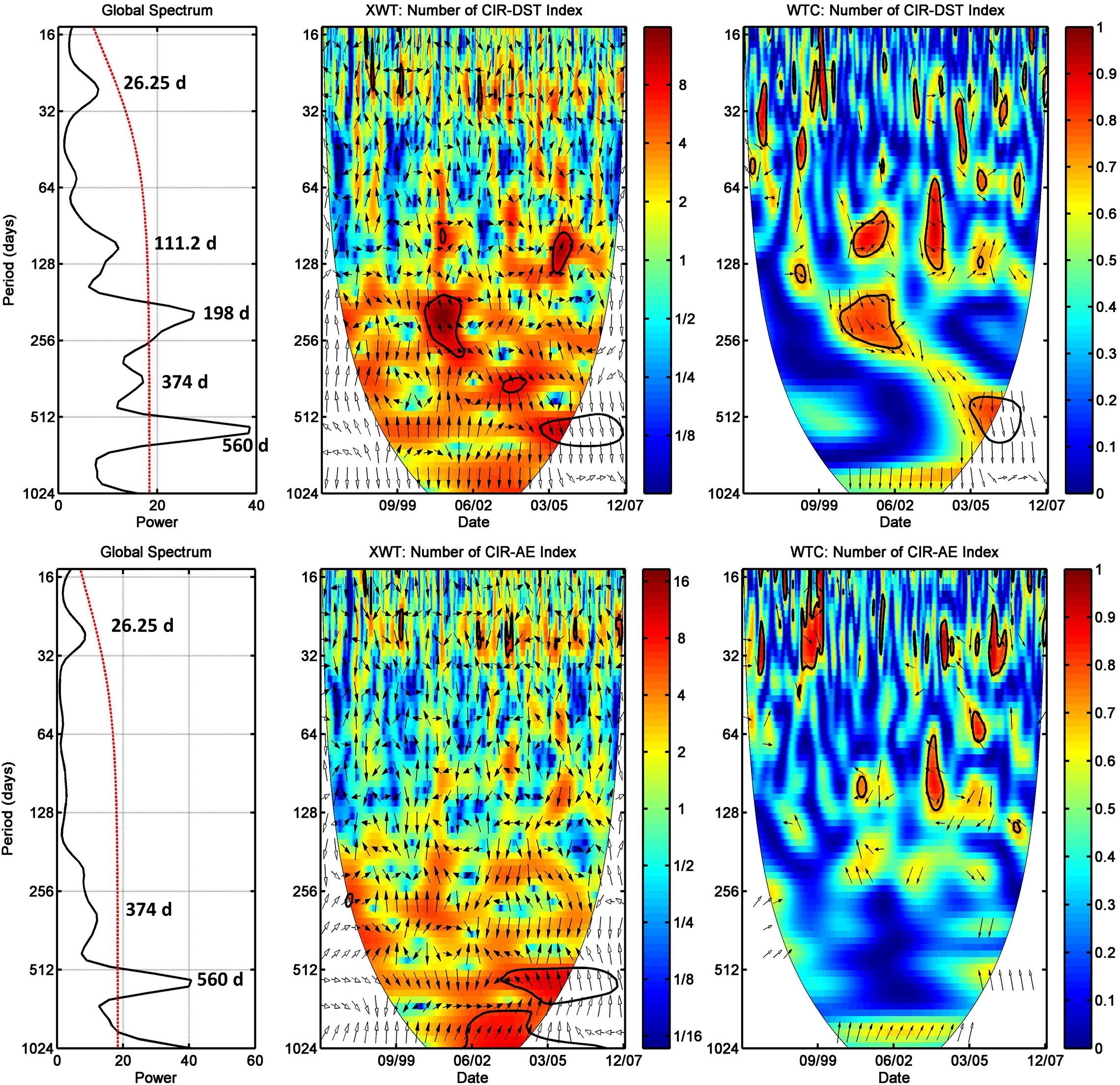} 
\caption{Cross-wavelet transformation (left) and Wavelet Coherence (right) of the CIR occurrence and the geomagnetic indices, same as Figure \ref{ICME_Indices}}
\label{CIR_Indices}
\end{figure*}
\begin{figure*} 
\centering 

\includegraphics[trim=02.0cm 0.0cm 02.0cm 0.0cm,clip,height=0.30\textheight,width=\textwidth]{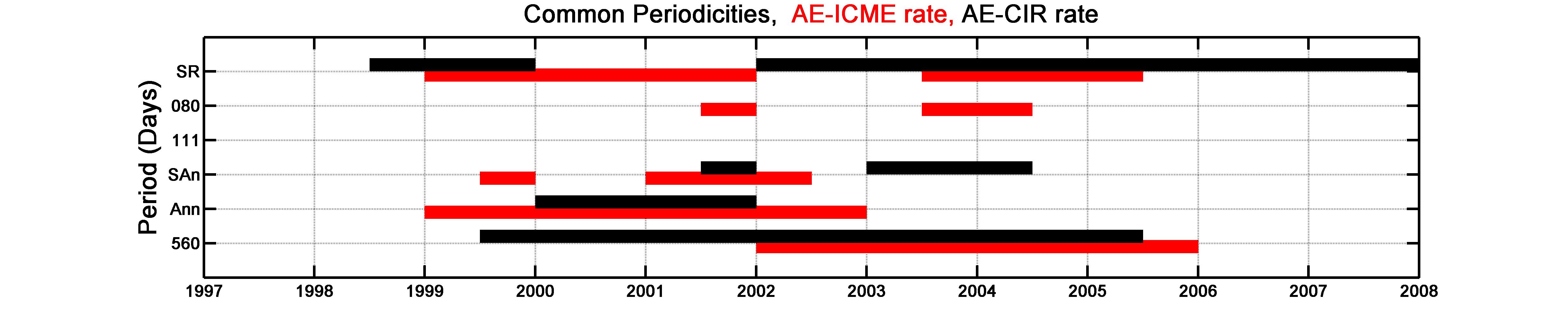} 
\includegraphics[trim=02.0cm 0.0cm 02.0cm 0.0cm,clip,height=0.30\textheight,width=\textwidth]{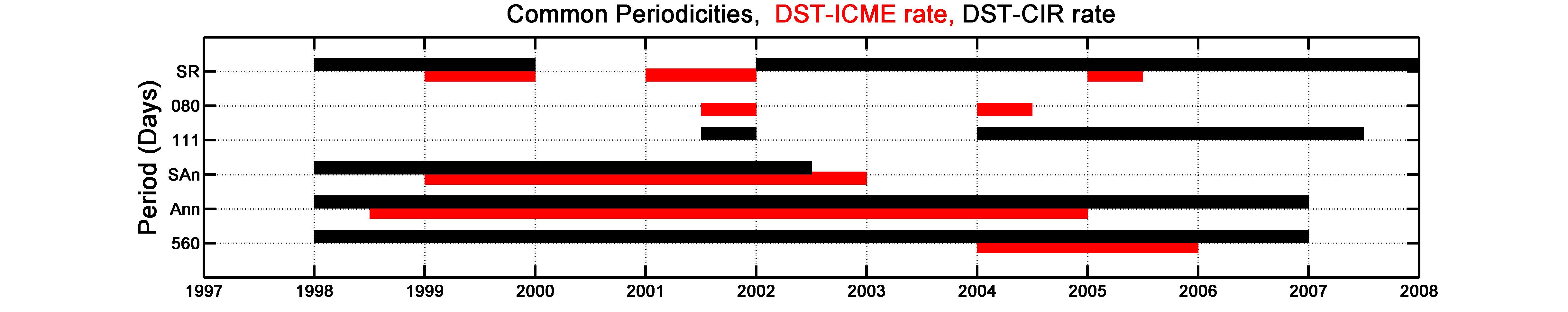} 
\caption{Common periodicities between geomagnetic indices (AE in upper panel and \DST~ in the lower panel) and drivers  as detected by the XWT. The red lines correspond to ICMEs and the black to CIR. The y-axis labels SR, SAn and Ann stand for Solar Rotation, Semi-Annual and Annual periodicities.}
\label{BAR_Indices}
\end{figure*}

\subsection{CIR--Geomagnetic Effects Relationship}\label{CIRGeo}
{ We examine the CIR rate relationship to the \DST ~and AE time series.  The results of the cross--wavelet transform (XWT) and wavelet coherence (WTC) are  presented in figure \ref{CIR_Indices}, in the same form as in section~ \ref{ICMEGeo},} and described below: 

\begin{itemize}

\item{ CIR--\DST~index: An 111 days peak in the decline phase exceeds the 95\% confidence level in 2005. There is an approximately semi-annual, 198 days peak that exceeds the 95\% confidence level in 2001-2002 but is also present in the rising and maximum phase of the cycle. The annual periodicity is present, with very low confidence level, in the entire cycle. In all cases the phase relationship is varying.}

\item{ CIR--AE index: The cross wavelet spectrum (XWT) of the CIR rate--AE index time series has a low confidence (less than 95\%) peak (374 days) during the maximum of the cycle (semi-annual periodicity was not detected).}

\item{ Intermittent, and, mostly low confidence (under 95\%) 27 days (ephemeral) peaks appear during the rising and decline phase of the cycle. This behavior is common to the ICME rate the \DST~and the AE, time series.}
\end{itemize}
 
{\noindent The 560 days periodicity appears in both CIR--\DST and CIR--AE XWT during the whole solar cycle yet is above the 95\% confidence level only during the decline phase.}

\subsection{Discussion} \label{Disc}

In this report two geomagnetic indices (daily values), each quantifying a different magnetospheric process (\DST~for ring current and AE for substorms) were examined. In place of the transient phenomena we used the daily number of ICMEs occurrence and for the recurrent phenomena the daily number of CIRs occurrence; these are the two drivers of the separate components.

The results in subsection \ref{wavelet} reveal short to medium periodicities in the range of days up to the year within the solar cycle 23. They are consistent with previous work by \citet{Katsavrias2012} where similar periodicities were detected within a sample spanning four solar cycles. In their report a number of spectral peaks where found with confidence exceeding 99\%. In this study, due to the smaller sample, most of these peaks appear below the 95\% limit. 

The ICME rate time-series has three significant components with periods of about 25, 66  and 187 days respectively; they are both quite pronounced during the rising phase and maximum of the cycle (see Fig. \ref{F1}). This result is, in part, consistent with the \mbox{$\approx$100-200} days periodicities {\cp[including the 153 day periodicity by~][]{Rieger1984}} of ICMEs per solar rotation reported by \citet{Richardson2005} and \citet{Richardson2010}, during the maximum and the decline phase \mbox{(2004-2005).} On the other hand, the CIR rate has  two pronounced frequency components of about 27 and 111 days in the decline phase of the cycle (see Fig. \ref{CIR}). The wavelet spectra of the geomagnetic indices \DST~ and AE show also the 27 days periodicity and, in addition,  a strong annual component (peak at 374 days) while the semi-annual variation appears pronounced only in the \DST~ index. The latter is probably the result of  the \ct{Russell1973} effect  which links the Earth's orbital position to the southward component of the interplanetary magnetic field; the ring current (and \DST), being associated to the dayside-reconnection which depends strongly on this component, is significantly affected. This is not the case with the AE index as this is  driven, mostly, by ram-pressure at the magnetotail (night-side reconnection) and is not as sensitive to the southward component.

Quite often the effects of these drivers are not distinct so the driver--component pairs are not easily separable in time, yet, by means of wavelet analysis (CWT, XWT, WTC) separation in the time--frequency plain may be obtained. Our examination indicates certain periods in time, or intervals in frequency (period) where some component becomes dominant for one or more indices.
 
The unusually active decline phase of solar cycle 23 \cp[see~][and references within]{Kossobokov2012} is an example of the importance of such time--frequency separability because, although the fast solar wind geomagnetic effects are dominant, there is also a significant contribution from transient flows (ICMEs) which originated from a higher--than--expected number of CMEs. Under normal circumstances the CIR-driven storms should generally occur in the the rise phase and then into the late declining phase of the solar cycle while the CME-driven storm should prevail at solar maximum \citep{Gonzalez1999, Yermolaev2002} as CME occurrence frequency and their velocity are both greatest during solar maximum \citep{Gopalswamy2004}. The separation between the two is, within the decline phase, only possible in the frequency (period) space as the shared short--term periodicities between ICMEs and geomagnetic indices and the shared short-term periodicities between CIRs and geomagnetic indices, both quite pronounced, do not overlap. 

The examination of the drivers--Geomagnetic Effects relationship, by means of cross wavelet spectra and wavelet coherence, in subsection \ref{ICMEGeo} and \ref{CIRGeo} are summarized in figure \ref{BAR_Indices}. The length of each bar in the chart represents the interval where common periodicities between driver and magnetospheric index are pronounced in the cross-power spectra (WXT). It indicates the following:

In the In the 27 day periodicities the CIRs modulate the two geomagnetic indices with interference from ICMEs during the cycle maximum and the extremely active period 2003; the  27 days CIR modulation of the geomagnetic indices does not appear at the cycle maximum where the major driver is the ICMEs. As regards annual and semi-annual periodicities (374 and 187 days) the ICME and CIR modulation of the geomagnetic disturbances overlap throughout the cycle 23. The 560 days periodicity, on the other hand, is dominated by CIRs during the whole cycle; the ICME contribution is localized within the unusually active decline phase 2002--2005(see Fig. \ref{BAR_Indices}).

The results, presented above, point to an anti-correlation of recurrent and transient phenomena in both geomagnetic indices as regards the 27 days component (see Fig. \ref{BAR_Indices}). This is consistent, in part, with the results of \citet{Feynman1982} and \citet{Du2011a} which demonstrated the anti-correlation of the recurrent and transient effects for a number of solar cycles using, however, low resolution data. For the component with period of \mbox{1.5} year the CIR seem to be the dominant driver again for both indices. For the remaining components, presented also in figure  \ref{BAR_Indices}, the driver--responce relationship is more complex and is not always the same for the \DST~ and AE. 

The common periodicities shared between ICMEs and the geomagnetic indices show prolonged periods of phase-locked behaviour (i.e consistent phase relationship) for components with periods of about a year. Furthermore, the ICME rate leads DST by 90\DG ~ which corresponds to a time lag of about 3 months whereas, in contrast, the ICME rate is in phase with AE index.  The physical origin of this intriguing difference between the phasing of Dst and AE relative to the ICME rate requires further investigation and is beyond the scope of this paper.

The common periodicities shared between CIRs-Geomagnetic indices show strong variations concerning the phase relationship as the CIR associated magnetic field, and the highly geoeffective z-component in particular, fluctuates strongly in a complex way. 

\section{Conclusions}\label{Conc}
{In the present study the relationship between transient (sporadic) and recurrent phenomena, ICMEs and CIRs, and the corresponding magnetospheric response represented by geomagnetic indices (\DST~ and AE) was examined. For the examination of this relationship between the drivers and the corresponding magnetospheric response we used the cross-wavelet transform (XWT), and wavelet coherence (WTC).  

Our results indicate that:

\begin{enumerate}

\item{CIRs modulate the geomagnetic responce during the rise and decline phase while ICMEs during the maximum of the cycle and the unusual active period of 2002--2005; the phase-relationship varies strongly in all cases for both drivers. Therefore there is an anti-correlation of recurrent and transient/sporadic phenomena throughout the solar cycle 23 but it is evident in the 27-days periodic component only.}

\item{The only clear phase-locked behaviour was found in the XWT of the \mbox{ICME-\DST}, ICME-AE components with periods of \mbox{$\approx$1.0} year. In the \mbox{ICME-\DST} case the phase difference corresponded to a time lag of about three months, while the ICME-AE XWT exhibited in-phase behaviour.}

\item{The component with period of \mbox{$\approx$1.3-1.7} years of the CIR time-series seem to be the dominant driver for both indices throughout the whole solar cycle 23.}

\end{enumerate}

A future study with a larger data--set (exceeding one solar cycle) is necessary in order to verify these results and expand in  larger time-scales.}


 \section{Acknowledgement}
The free for non-profit use MATLAB package of the National Oceanography Centre, Liverpool, UK was used in this work. The authors thank S. Patsourakos and the anomymous referees for helpful suggestions.
 
\section{References}
%


\end{document}